Draft chapter for: Tara Behrend (Ed.), *Human-Technology Partnerships at Work.* Cambridge University Press.

# Implementing New Technology in Educational Systems

Scott Allen, Lisa Bardach, Jamie Jirout, Allyson Mackey, Dana McCoy, Luca Maria Pesando, and René Kizilcec

## Educators and Technology in Education

Educators are more than workers within educational systems; they are stewards of educational systems. Their job is not only to deliver lessons, grade assignments, provide feedback, and meet with parents; they must also analyze student performance data, identify patterns that inform targeted interventions and personalized learning plans, continuously develop the curriculum, set ambitious learning goals and use up-to-date pedagogical theory to adapt instructional strategies, act as advocates for educational policies that promote inclusivity and equity, and much more (Reich, 2020). Most educators deeply care about the learning and wellbeing of their students and colleagues. Given the chance, they will do whatever they can to make improvements to these ends. In this role as architects of change, educators deal with conflicting definitions of success, multiple stakeholders, complex causal relationships, ambiguous data, and intricate human factors. Amid all this, most educators and the educational systems around them are strained to the capacity of what their time, training, and budgets allow. The problem is not merely that they must perform demanding tasks, but more so that they must constantly implement improvements and interventions amid the complex challenges of the *organizations* in which they work. These challenges can be especially difficult in implementation of related education technology, which is continuously developing at sometimes rapid pace (Ross, 2020). Whether the context is an individual classroom, a school district, or a postsecondary institution, implementing beneficial human-technology partnerships requires attending to the needs and constraints of these classrooms, districts, institutions, and so forth as organizations and engaging in this work as a partnership with educators. This chapter lays out the principles and processes of developing successful educator-technology partnerships including key considerations for each step and an example protocol for engaging in this endeavor.

New technologies are often proposed as solutions to improve educational systems (or even "revolutionize" education itself), but attempts to impose technological solutions often fail spectacularly (Daniel, 2012; Cuban, 2003). In 2013, the Los Angeles Unified School district excitedly budgeted over $1 billion to provide every student, teacher, and administrator with an iPad without first considering what needs these devices were supposed to meet. Adding technology for technology's sake and attempting to make these devices the centerpiece of

day-to-day operations was not only unnecessary, but outright disruptive to students' learning. Without a discernible need for the new devices, many teachers coped with the disruption by simply not using them ([Lin, 2015](); [Lapowski, 2015]()). Even when new technology is designed to meet educational needs, technology developers do not always consider the constraints present in the environments where it is meant to be used. In 2005, driven by the belief that access to technology could empower low-income communities around the globe, a team of influential innovators, educators, and technologists hatched an ambitious (now infamous) plan to provide One Laptop Per Child everywhere ([Kraemer et al., 2009]()). To their credit, they performed small-scale pilot tests of the supposedly rugged, low-cost laptops they had developed. These pilots revealed the need to revise the initial idea behind the project due to lack of sufficient tech support and infrastructure in the target communities. Rather than collaborating with members of these communities to revise their idea, leaders of the project marketed the idea to Western philanthropists by telling heartwarming tales of children who used the laptops to learn to read. In the end, simply building libraries in the target communities would likely have done more good than designing and distributing special rugged laptops.

Although certain burdensome tasks performed by educators (tedious grading, feedback, communication, bookkeeping, etc.) are reasonable opportunities for technological solutions, technology developers must change the way they define these practical problems faced by educators. Too often, technology developers erroneously suppose they are solving *well-structured* problems. A well-structured problem is one with a clear definition of success, unambiguous data, and simple causal relationships ([Simon, 1973]()). It seems straightforward enough to reason that if giving detailed personal feedback to students is too time consuming, an AI chatbot that automates this task solves the problem. The key point of failure is not the capabilities of the technology itself, but the process of incorporating that technology into complex educational systems ([Reich, 2020](); [Ames, 2019](); [Dancy & Henderson, 2010](); [Allen & Kizilcec, 2023](); [Wieman, 2017](); [Damshroder et al., 2022](); [Cuban, 2003]()). This should not be too surprising to anyone who has observed the challenges of incorporating new technology, or any new work practices, in their own workplace.

Even when the technology works as promised and the problem it is intended to solve is a well-structured one, incorporating that solution into an educational organization is a remarkably *ill-structured* problem ([Dancy & Henderson, 2010](); [Damshroder et al., 2022]()). Ill-structured problems have many paths to various solutions and conflicting definitions of success ([Simon, 1973]()). The distinction between well-structured and ill-structured problems is, interestingly enough, a significant concept in the fields related to machine learning as well as human learning. The term "ill-structured problem" is credited to Herbert Simon ([Simon, 1973]()), who is highly respected in pedagogy and computer science communities alike. Most readers will not need to review the works of Herbert Simon to understand how to overcome the challenges of making improvements in educational organizations. You probably already know how to solve, and have experience solving ill-structured problems within the scope of your profession. Think of a time when you were solving a problem, and it was not clear initially what problem you were even solving. One with multiple possible paths to various solutions that may impact multiple people in differing ways. How did you approach this problem compared to one with a clear path to a solution? You probably used some of the following strategies: You collaborated with multiple people, including those who would be impacted by the outcome, and those who may have insight into possible solutions. You set flexible timelines, maintained active channels of communication, and carefully monitored progress along the way. You made a plan that balances and prioritizes the most important needs and constraints of the people involved. In short, you used these strategies because you knew to anticipate revision. When working on a well-structured problem without the need to anticipate revision, you simply work to build a solution; but when working on an ill-structured problem, you use these strategies to build a team, build a theory, and build a solution.

All of these strategies are strongly supported by evidence from Implementation Science, which is the scientific study of incorporating new technology into organizations (Damschroder et al., 2022; Moulin et al., 2020). The core principles and processes of Implementation Science provide a framework to approach these problems more like an expert designer who spends time and resources to fully understand the problem, develops and tests a predictive model for solving the problem, and iteratively adapts to new complications ([Perry et al., 2019](); [Moulin et al., 2019]()). After deeply exploring stakeholders' needs and multiple solutions to meet those needs, this implementation process provides resilience to challenges and uncertainties that will arise by using incremental testing to expose unforeseen challenges, collaborating with local leaders to adapt accordingly, and assimilating the new intervention into the regular operations of the organization. The framework of these processes provide the opportunity to engage in expert-like design thinking, but is not so formulaic to allow for a perfunctory approach. Successful educator-technology partnerships require thoughtful application of principles of thorough inquiry, teamwork and communication, and iterative revision.

# Principles for Developing Successful Educator-Technology Partnerships

Ill-structured problems demand a progression of first deeply exploring the nature of the problem followed by iteratively testing possible solutions. In the case of implementing new human-technology partnerships, there is an additional end phase of solidifying that new partnership for the long-term. Though each phase of implementing a new technology has a unique focus, and each setting has unique needs and constraints, a common set of core principles is essential throughout. From the very beginning, the principle of thorough inquiry grounds the project in a full understanding of local needs, the principle of teamwork and communication connects the project to the source of that understanding, and the principle of iterative revision initiates the project's adaptation to the unique local context. As the plan advances to small-scale testing, these same three principles provide resilience to setbacks, and allow the project to capitalize on successes. Even once the final version of the intervention is being used at full-scale, it is not fully implemented until it is subject to the ordinary ongoing revision practices of the organization ([Moulin et al., 2020]()). All three principles are mutually complementary. Inquiry and collaboration are means to the end of making necessary revisions, collaborative communication should always be done with a mindset of genuine inquiry, and any inquiries or revisions should be done in full collaboration with teachers and students.

## Thorough Inquiry

Thorough inquiry occurs at all phases of a successful implementation due to the ill-structured nature of implementing new educator-technology partnerships, the high likelihood of setbacks, and the inevitable need for revision ([Moulin et al., 2019](#)). The first opportunity for thorough inquiry is a needs assessment process where the needs of the students, teachers, and organization as a whole are considered *before* an intervention is selected to meet those needs. A simple example from instructional design is outlining learning objectives before selecting course content and teaching methods. Without a thorough needs assessment, an implementation can fail simply because it does not solve any problems or provide anything of value. Avoid post-hoc justification of a pre-selected intervention. Collaborating with local stakeholders during the needs assessment is essential because the needs of the organization depend on local contextual factors, and different levels within the organization can have unique needs. This collaborative process will likely highlight unique elements of the organizational structure and its interaction with the external environment. Thus, a rigorous needs assessment will influence the teamwork and communication strategies to be used throughout the project.

Beyond the initial needs assessment, there are many contextual factors to be explored and mapped out in detail ([Damshroder et al., 2022](#)). This includes (a) the features of the new intervention, (b) the features of the organization, and (c) the environment in which the organization operates. There are many frameworks for itemizing all the contextual factors, but no proven formula for then developing an implementation plan. Thus, it is essential to create a detailed map of the interconnections among these features ([Powell et al., 2015](#); [Damshroder et al., 2022](#)). Mapping out the interconnected factors in this way is the *Theory of Change* approach ([Bruer et al., 2016](#)). This carefully mapped out Theory of Change (ToC) generates a set of *many* possible implementation plans and judgments about which factors are most influential. For example, introducing iClickers to enable interactive engagement in a university lecture setting could influence students' end-of-semester feedback. Student feedback can influence tenure and promotion decisions, which can influence instructors' willingness to use the new technology. Deeply exploring these interconnections before foreseeable problems arise leads to much stronger implementation plans.

As a new intervention is incrementally tested and monitored, there are opportunities for both exploratory and confirmatory measurements. Consider implementing iClickers for interactive engagement in lecture, but in a particular context, such as a cultural context where students are accustomed to silently listening to lecturers rather than interacting during class time. Since the instructional methodology (interactive engagement) has been proven to yield specific positive effects on student's learning ([Freeman et al., 2014](#); [Lovett et al., 2023](#); [Schwartz et al., 2016](#)), it is appropriate to use confirmatory measurements to determine whether the new intervention is producing these benefits as expected. Often these confirmatory measurements will be quantitative, such as students' test scores. But since the interaction of this methodology with the unique cultural and organizational context is unknown, and thus not all of the effects can be anticipated, exploratory measurements are also appropriate. These exploratory measurements are often qualitative, such as student feedback forms. One type of qualitative exploratory measurement that tends to be overlooked is sitting in on classes and observing how students and teachers are interacting with the new technology. Overall, the process of incrementally rolling out a new intervention will go smoother if the main effects of the intervention and its complex interactions with the specific application context are fully explored in the early planning stages. And yet, because unexpected events are inevitable, thorough inquiry is still necessary even in the middle of a project.

In the later stages, an intervention that produces the intended effects is not guaranteed to last. For example, implementing online learning technology in traditional in-person classes has allowed many university courses to deliver the bulk of the course content online and refocus instructor-student contact time on coaching students rather than lecturing them. In many STEM courses, the existing classrooms for these engaging coaching sessions are large lecture halls. This physical infrastructure served as such a strong *stability mechanism* for old lecture practices, that many courses naturally reverted back. Every organization has stability mechanisms that can present a barrier, or can be leveraged for long-term success. Carefully consider what has stayed the same for a long time within the school (e.g., scheduling processes), and what forces are imposing on that stability (e.g., remote work and learning). Another feature which yields long-term success are the school's existing quality assurance processes. Inquire about what the school already does well, and what processes have been in place to maintain that quality. For example, one university engineering department may consistently maintain up-to-date learning objectives and appropriately high grading standards as a result of having processes in place to take in feedback from firms that hire their students. If the department slips in adequately preparing their students, feedback from those firms will reliably initiate corrective updates to the course learning objectives. Another engineering department may meticulously avoid unfairly failing any students because the department has practices in place to listen to students (or students' parents) who raise concerns about fair grading. If students start getting unfairly low outcomes, the resulting complaints will reliably initiate corrective actions in grading practices. Educational organizations of all kinds have their own unique quality assurance practices. Once the implementation project is over, the existing stability mechanisms and quality assurance process will determine whether the new intervention persists. By knowing what has stayed the same, what is done very well, and why, existing stability and quality assurance mechanisms can be applied to sustain the long-term success of the new intervention. This degree of careful, thorough inquiry requires allocation of time and resources in the initial exploration and preparation phases. It also demands time and resources for monitoring, observation, and collaboration in later phases.

## Teamwork and Communication

The Los Angeles Unified School district set themselves up for failure when they began the project by talking to high-profile technology vendors about the products they wanted. The year before that, the leadership of Milpitas School District (a nearby district with similar challenges and goals) set themselves up for success by talking to teachers and local principals first. They asked them "If you could design the school of the future, what would that look like?" ([Lin, 2015](#);

Lapowski, 2015). This provocative question was not the opening line of a sales pitch; it was a genuine invitation for local schools to create a shared vision with district leadership. Together, they implemented technology into the classroom that actually solved many of their problems while integrating smoothly with day-to-day operations. The successful human-technology partnership implemented by the Milpitas School District was made possible by building essential human-human partnerships.

Strong implementation plans are built in collaboration with all levels of an academic organization (Daniel, 2012; Reich, 2020). Students and teachers are the best source of information about the needs and constraints of students and teachers. They are not merely clients, they are essential partners in the implementation process. In many contexts school leadership plays a key role in driving and sustaining the change. Erroneous assumptions about the role of leaders within the local context can be detrimental. High-level administrators, department heads, instructors, teaching assistants, and students have their unique roles and are subject to unique influences. These roles and influences come to light by engaging in meaningful collaboration with all levels of the academic organization. People will also take on non-formal roles within the organization, from champions who will enthusiastically support the change, to naysayers who will push back against it (Dancy & Henderson, 2010; Wieman, 2017). They play an important role in conducting the needs assessment, selecting an intervention, developing goals and timelines, recovering from inevitable setbacks, and sustaining the change long-term. A strong implementation plan reflects the collaborative input of these informal roles as well.

This is not to say that a bottom-up, grass roots movement by students and teachers is the only starting point for success. After all, the human-technology partnerships implemented by the Milpitas School District was initiated and led by the district superintendent, but they intentionally led it in such a way that local schools were creating the central goals of the project. Successful implementations have many possible starting points so long as they quickly foster meaningful involvement from all levels. Projects initiated from the top-down or initiated externally do have a notable disadvantage which is that they don't spontaneously yield teamwork and collaboration among students and teachers within the organization. Teachers are far more likely to overcome frustrating setbacks with new interventions when they can talk about it with other users. In many cases, the single biggest factor in teachers overcoming setbacks is substantive support from leadership (Dancy & Henderson, 2010). Regardless of the starting point, do not delay in establishing these human-human partnerships. By collaborating early, and actively maintaining open lines of communication, projects to integrate new technology into educational systems will be more adaptive to early setbacks and more prepared to capitalize on early successes.

## Iterative Revision

Each phase of an implementation project is an iterative process because the project must not advance from one phase to the next until critical milestones are achieved. The LAUSD advanced to purchasing iPads before achieving the milestone of selecting an intervention that met local needs. They then went on to advance to full-scale implementation of those devices in every classroom before achieving the milestone of identifying the technical and organizational challenges that the full-scale implementation would entail. Table 5 provides an itemized list of each milestone that marks readiness to advance from each phase to the next. Approaching each phase as a set of processes to iterate upon (rather than just a linear sequence of steps) allows the project to set high standards in achieving these milestones. If only one or two rounds of the processes in the initial exploration phase has not yet identified an intervention that meets local needs to a high standard, iterate on those processes again until it does. Likewise with each subsequent phase: iterate within that phase until the milestones are achieved to a high standard.

Along with a readiness to iterate the processes within phase, the project should maintain a readiness to make methodical revisions to the products of each phase (the team, the theory, and the solution). The OLPC project had a clear opportunity to revise their idea and did not take it. Initial production runs of the cheap, rugged laptops revealed that they were not so cheap; and initial tests revealed that they were not so rugged. More importantly, the laptops did not have nearly the expected impact on the communities in which they were intended to empower. As difficult as it must have been for the OLPC to confront a fatal flaw in their central idea, these tests revealed that laptops were simply insufficient to empower low-income communities around the globe. If the initial formation of their idea had been done in collaboration with those communities, there may not have been a need to make such difficult revisions. Nonetheless, the need for revision arose and the plan rolled forward unrevised. Even in properly planned projects, unexpected challenges will arise and unrevised plans rarely succeed ([Reich, 2020](#)). The nature of ill-structured problems is that it is not possible to anticipate every complication. Yet it is possible, and essential, to craft a strategy whereby unforeseen complications generate meaningful revisions. Plans which allow for only minor revisions of the intervention are not sufficiently flexible. Every aspect of the plan should be open to multiple major revisions. The goals, timelines, evaluation processes, team member roles, collaboration approaches, and resource allocation must all be allowed to adapt to new information and changing conditions on the ground. The Theory of Change is both a basis for seeking these revisions and is itself subject to revision in response to ongoing communication with stakeholders and monitoring of implementation progress. As the following section describes how to conduct a successful implementation, remember that it is not the brilliance of the initial idea that leads to success, but methodical (and sometimes difficult) revision. One of the central purposes of these processes is to enable those revisions.

# Processes for Developing Successful Educator-Technology Partnerships

The long-term success of a new technology, or any new practice, does not merely depend on it working as promised, but the process by which it is implemented ([Aarons et al., 2011](#)). To provide a process which fully utilizes the three core principles and avoids the common pitfalls, we offer two complementary planning tools: the EPIS Framework (Exploration, Preparation, Implementation, and Stabilization) from Implementation Science ([Aarons et al., 2011](#)); and SWOT Analysis (Strengths, Weaknesses, Opportunities, and Threats) from Project Planning Research ([Madsen, 2016](#); [Namugenyi et al., 2019](#)). The EPIS Framework outlines a process for planning and conducting an implementation project such that no critical steps in the process are skipped (Moulin et al., 2019), while SWOT analysis stimulates creative problem-solving in selecting strategies to overcome the unique challenges and opportunities present in the local context of a project. This section describes the processes step-by-step. Figure 1 shows *what* the project is producing: a team to collaborate together, a Theory of Change to understand the problem, and a solution to that problem. Tables 1-4 describe *how* to plan and conduct the project: the processes in each phase, how the core principles are manifested in each phase, and the pitfalls to avoid. The final section will provide *protocols* for using the EPIS framework: milestones for determining when to advance from each phase the next, and instructions for using SWOT analysis at the points where creative brainstorming is appropriate.

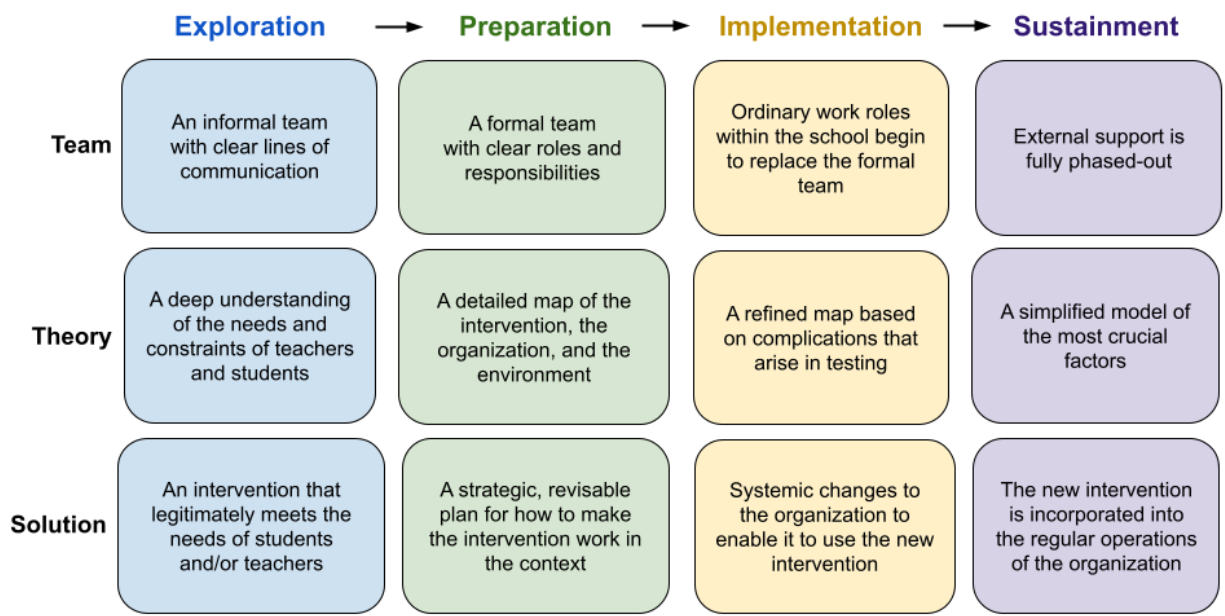

**Figure 1**. The evolution of the team, the theory, and the solution through the four phases of a project to implement new human-technology partnerships in education.

The process is divided into distinct phases because the end results of each phase lay the groundwork for solving the problems in the next. A successful educator-technology partnership must be founded on a meticulous exploration of the needs of local stakeholders (primarily students and teachers), multiple possible solutions to meet those needs, and whether these solutions fit the constraints of the local context. Only after finding an intervention that legitimately meets local needs and fits the constraints is it possible to prepare to implement that intervention. Implementing a new technology or a new practice into any type of organization can strain the capacity of that organization. This is particularly important to remember when that organization is a school with an already strained budget and workforce. Even in preliminary planning, this requires thinking ahead about how to build the capacity of the organization, how to offer any needed external support, and how to sustain the intervention once that external support is gone. These plans will almost certainly be modified once small-scale tests reveal unforeseen challenges and early successes. Thus, each phase of the EPIS process depicted in Figure 1 focuses on a distinct two-fold problem which must be solved in preparation for the subsequent phase.

The problem of the exploration phase is to fully understand the needs and constraints of local stakeholders, and to select the correct intervention. Rushing through this problem haphazardly will leave a weak foundation for the next phase. As the project advances from the exploration phase to the preparation phase, the problem shifts to fully understanding the intricacies of the specific educational context, and to figure out how to make the chosen intervention work as expected amid those intricacies. This problem is not fully solved until it produces a plan that is highly adaptable in the face of the problem in the next phase. The problem in the implementation phase is how to interpret the complications that will arise, and which parts of the plan to revise. In the sustainment phase, it will be difficult to solve the problem of cementing the new intervention and phasing out external support if the solution is still not working or causing strain on the organization ([Aarons et al., 2011](#)). In the remainder of this section, we explain the essential subprocesses of each phase, how the three core principles are manifested in each, and common pitfalls to avoid. In the next section, we describe the milestones marking the successful completion of each phase, and the ideal mindset to maintain.

## Key Considerations in the Exploration Phase

Before exploring potential interventions, it is essential to fully explore the needs of local stakeholders[1]: educators, school leaders, parents, and especially students. This helps to avoid making a post-hoc justification of a preselected intervention. Table 1 shows important processes, principles, and potential pitfalls to consider in the exploration phase of a new technology intervention in education.

**Table 1**. Processes, principles, and pitfalls in the exploration phase of developing an educator-technology partnership.

| Exploration Phase | | |
|---|---|---|
| **Processes** | **Principles** | **Pitfalls** |
| Needs Assessment<br>Develop a full understanding of local needs before selecting an intervention.<br><br>Intervention Fit Assessment<br>Consider several solutions to meet local needs, and scrutinize the compatibility of each need to the local | Thorough Inquiry<br>Explore a wide variety of local needs and potential interventions.<br><br>Teamwork and Communication<br>Stakeholders within the academic organization take an active role in the early planning, especially needs | Premature Commitment to Initial Ideas<br><br>Failure to Consider Constraints<br><br>Insufficient Stakeholder Collaboration |

---

[1] While the term "stakeholder" has outdated connotations in some contexts, in the context of Implementation Science it is widely used to refer to community members who engage in meaningful partnerships while respecting the variety of intersecting stakes they experience.

| context. | assessment and fit assessment. | |
| Preliminary Revisions Explore several possible adaptations of the intervention to better meet the needs and fit the context. | Iterative Revision Continually re-examine local needs while exploring and adapting possible interventions. | |

## Needs Assessment

Each educational system may have vastly different needs. A detailed understanding of stakeholders' most important needs must precede the selection of an intervention to implement. The depth and formality of the assessment will depend on the scale of the changes implemented. When the United States Navy was preparing for a major curriculum update to their training program for nuclear operators, they first spent several years conducting a multimillion-dollar study itemizing hundreds of decisions that a nuclear operator in the fleet must make. This comprehensive catalog of decisions allowed the Navy to identify all the skills which must be included in the new curriculum. In contrast, when a small working group of instructors from the initial classroom phase of that same training program was planning a project to improve their classroom teaching methodology, they spent several months doing interviews and focus groups with experienced nuclear operators regarding their thought processes during fleet operations. This qualitative understanding of experienced operators' thought processes helped the working group identify when to use certain teaching techniques optimized for learning memorized procedures versus techniques optimized for adaptive decision-making. At the scale of a single classroom, the assessment might take the form of informal conversations between teacher and students, and perhaps a brief survey.

In any educational setting, the needs will often focus on student learning outcomes, but it is important to consider a broad scope of needs. In addition to students, other stakeholders of a school organization strongly influence student outcomes. Teachers, administration, and other stakeholders such as parents, play distinct roles in helping students thrive. At the university level, potential employers often have a valuable perspective regarding the challenges new graduates will be expected to face. When conducting surveys, focus groups, and interviews, be sure to include all kinds of stakeholders connected to students' needs. These essential roles will also vary from school to school. Certain needs that are vital to stakeholders in one school may be secondary in another school. However, no school is so unique that one must reinvent the wheel to meet their needs. There will be existing peer-reviewed literature to help better understand the needs that come up during this assessment, and to better understand possible solutions.  Some needs may be in direct conflict with others. Thus, explicitly prioritizing those needs is a difficult but absolutely necessary component of the assessment process. Understanding the needs of local stakeholders is inherently linked to understanding the organizational structure and the specific roles of organizational members. The most common and detrimental pitfall is to skip this process or to conduct it with insufficient rigor due to overconfidence in the initial solution. Avoid simplistic pictures of student needs and the needs of other stakeholders. Diligence and rigor in assessing these needs establishes the foundation for a successful implementation project.

## Intervention Fit Assessment

The most crucial decision of the entire project is to select an intervention that legitimately meets the highest priority needs of students and schools. There are likely many possible mechanisms to meet these needs. The intervention fit assessment is, in essence, a brainstorming process. As with all brainstorming processes, it is best practice to start by prioritizing quantity over quality of ideas. exploring many ideas of varying quality, later revise and edit. Another best practice for brainstorming is to stay focused on a central issue or question. Thus, if the top priority needs are still unclear, it is better to backtrack to the needs assessment than to blindly press forward. Conjecturing possible solutions can expose previously unconsidered needs. After exploring solutions, explore constraints. Consider if there are any local contextual factors which may enable or inhibit each possible solution. An intervention which is successful in one school may be inappropriate or impractical in another. Temporarily resist the urge to rule out impractical ideas. When exploring constraints, it is easy to slip into the mindset of editing out ideas rather than brainstorming. Continue with the mantra of quantity over quality. Itemizing potential failure mechanisms of the impractical ideas will stimulate a deeper exploration of the practical ones. Conducting a literature review can expedite this exploration. Even with somewhat new technologies, there is likely to be useful guidance in peer-reviewed research on implementing similar interventions in similar contexts. A word of warning: this exploration of contextual constraints must not be a post-hoc justification of a pre-selected intervention. Not only does post-hoc justification of a pre-selected intervention usually indicate a poor choice of intervention, but it also undermines the essential problem-solving processes of the next phase. Approaching the intervention fit assessment (and the exploration phase overall) by legitimately questioning which intervention is best will lead to a much deeper understanding of the important contextual constraints. These constraints will be analyzed in greater detail in the next phase, and additional constraints and enabling factors are likely to be revealed in the early stages of implementation. Moreover, plans which have explored only a few possible solutions are likely to be fragile and sensitive to minor revisions. Plans which have explored many possible solutions are more likely to have identified an intervention which is appropriate for the local context, and are more likely to have sufficient adaptability. The process of outlining these interconnected factors can be aided by using SWOT analysis, which we describe in the following section.

## Preliminary Revisions

Following the needs and intervention fit assessments, now is the time to edit ideas. Rather than simply crossing off solutions with obvious failure mechanisms, instead ask how each solution needs to be modified in order to be practical and effective in the local context. Since the selected intervention will typically need to be adapted to the local context, narrow down the choices based on the feasibility of the adaptation and the priority of the needs that it meets. Whenever possible, identify the causal mechanism by which the intervention achieves the intended outcome. It defeats the purpose to adapt the intervention to fit the context, but in so doing remove the causal mechanism that makes it work. Do not attempt to plan every revision in advance. Later phases will include small-scale tests of the new intervention, which will highlight the need for certain revisions, but it is also a mistake to wait to begin considering revisions. Planning ahead for multiple possible revisions can assist in developing a sufficient understanding of the local organization, developing an approach to monitoring and measurement, and preventing premature commitment to initial ideas. The level of detail when mapping out these interconnected factors will strongly influence the degree of adaptability in the face of setbacks. Exploring these revisions will help narrow down the possible solutions, but it will also highlight some gaps in the needs assessment. Never rush to decide on the single best option. Rather, iterate on the processes of assessing needs, assessing the fit, and exploring revision to identify the best solution.

## Key Considerations in the Preparation Phase

Significant strategic planning is required to manage the complexity of implementing new technologies and associated practices into schools. Without careful planning, even a highly effective intervention will likely fail to have the expected impact. Once an intervention is selected, plans for implementing that intervention are strategized around an in-depth analysis of many possible barriers and facilitators within the school system and school environment. Strong implementation plans are developed in collaboration with local stakeholders from the beginning and throughout the entire process. Unforeseen setbacks and complications are likely to arise and adaptability to these complications requires continuous monitoring and iterative revision of the implementation plan. The plan is ready when it enables the team to adapt to the setbacks, uncertainties, and complexities of the upcoming implementation process. Table 2 shows important processes, principles, and potential pitfalls to consider in the preparation phase of a new technology intervention in education.

**Table 2**. Processes, principles, and pitfalls in the preparation phase of developing an educator-technology partnership.

| Preparation Phase | | |
|---|---|---|
| **Processes** | **Principles** | **Pitfalls** |
| Formal Team Building
Establish a team to manage the implementation project consisting of external experts and local members of the school organization. Outline clear roles and establish lines of communication.

Organizational Capacity and Barriers Assessment
Outline how the capacity of the school organization (budgets, infrastructure, | Thorough Inquiry
A wide variety of local contextual factors are itemized in order to determine the strongest influences on individual and organizational behavior. Organizational capacity and organizational barriers are mapped-out in detail. Lessons learned are gleaned from any similar interventions in similar contexts. | Failure to utilize local stakeholder's knowledge of organizational capacity and barriers

Failure to collaborate with all levels of the organization

Insufficient resources allocated to team building and planning

Rigid timelines which do not account for inevitable |

| workforce, expertise, etc.) may help or hinder the success of the intervention and of the implementation project.<br><br>Implementation and Sustainment Planning<br>Craft plans to build organizational capacity, adapt to the unknown setbacks of the implementation process, and ensure the long-term sustainment of the intervention. Select measurement methods to observe the effects of the intervention. | Teamwork and Communication<br>A special team within the organization is formed to manage the implementation. Local stakeholders take an active role in the capacity and barriers assessments, and in the implementation and sustainment planning.<br><br>Iterative Revision<br>The intervention and implementation plan are revised in light of organizational capacity and barriers. Multiple possible future revisions are planned for potential issues arising in the implementation and sustainment phases. | setbacks |
|---|---|---|

## Formal Team Building

Decisions about the project are made as a team. Local organizational members (teachers, administrators, etc.) are the best resource for understanding the local context, and they must eventually take sole responsibility for sustaining a high-quality intervention in the long term. They form a team with external experts to manage the implementation process. In higher education, these formal teams are sometimes called Departmental Action Teams ([Corbo et al., 2015](#)). While team member roles may adapt throughout the process, there should be no ambiguity as to what those roles are. In particular, team members and local stakeholders know what power they have to make decisions that impact the progress of the implementation. In many organizations, leaders have a great deal of influence, but the specifics of their role will vary from one context to another. Just as local team members should clearly know their role in the implementation process, external experts need to have a clear and detailed understanding of the various roles within the regular operations of the organization. For the implementation plan to have sufficient adaptability, it should be derived from a map of the organization which accounts for all levels of organizational structure and the external environment in which the organization operates. Thus, organizational members at all levels must actively contribute to developing this map and using it to derive an implementation plan. As with the needs assessment process, the level of formality depends on scale. Formal teams will be much less formal at the scale of a single classroom.

## Organizational Capacity and Barriers Assessment

A new intervention and the process of implementing it may strain the organizational capacity of an institution or classroom. For example, a school may not have sufficient workforce, training, or technological capacity to utilize the intervention. If so, resources should be applied to appropriately boost organizational capacity or offset the intervention by reducing other strains on capacity. Each organization may have unique advantages, resources, or capacities which should be utilized throughout the implementation process. Moreover, each organization can face unique barriers to a particular intervention and to change itself. While some of these barriers will arise during early implementation, early consideration of these barriers can save time and resources. For an intervention involving a new teaching methodology, insufficient teacher experience with the methodology can be a barrier. For a technological intervention, the available technological infrastructure can be a barrier. If the intervention requires a budget to sustain it, there can be financial barriers. Various barriers can interfere with interventions of different kinds. Like the intervention fit assessment in the previous phase, SWOT analysis is an ideal tool for brainstorming and organizing all of these important factors. The results of this assessment combined with the results of the two assessment processes before the exploration phase (the needs assessment, and the intervention fit assessment) can be combined together to craft a Theory of Change ([Breuer et al., 2016](#)). If the Theory of Change is sufficiently detailed, it can be used for generating and revising plans for the later stages of the project. The Theory of Change is also subject to revision in light of results from small-scale tests.

## Implementation and Sustainment Planning

It is at this point that the intensive strategizing begins. Previously, most of the work was centered around building a full understanding of the problem at hand. Now, much of the work shifts to developing solutions to that problem. This is where the ill-structured nature of implementing human-technology partnerships in education must not be underestimated. No matter how much rigor has been applied to outlining all the contextual factors, there is no formula for strategizing what to do about them ([Perry et al., 2019](#)). Thus, it is still a process of creative brainstorming. As is the case with other ill-structured problems, there will be many possible solution paths. Diligence in using SWOT analysis to craft a detailed Theory of Change will pay off here. While we cannot provide a formula for making a plan, we can describe some key features of strong plans. A strong implementation plan must account for the later phases from the beginning. It must include an approach to incremental implementation, measuring relevant effects, monitoring progress, and revising accordingly. It must allocate time and resources to early testing phases. It must account for how quality assurance of the implementation will be managed long-term, and which stability mechanisms exist within the organization. Flexible plans with multiple possible approaches and moveable timelines are preferred to rigid plans with a single approach and fixed timelines. It is much more difficult to begin to leverage stability mechanisms after the intervention has been implemented at full scale. Thus, avoid the pitfall of waiting until the later stages to begin strategizing for long-term stability. As always, proposing solutions and making concrete plans may highlight gaps in the team or gaps in the theory. Perhaps the team may require some outside expertise, or additional resources for communication and collaboration. Perhaps the Theory of Change is missing some important nuance. Iterate on building a team, building a theory, and building a solution until the project is ready to move ahead to implementation.

# Key Considerations in the Implementation Phase

In each of the first two phases, it is usually advantageous to iterate on the processes therein until the project is ready to move on to the next phase. The implementation phase is always inherently iterative. Adequate planning in the preceding phases will not fully guarantee that the implementation phase will go smoothly. Rather, the implementation phase should begin with a readiness to adapt to the uncertainties and challenges to come. Table 3 shows important processes, principles, and potential pitfalls to consider in the implementation phase of a new technology intervention in education.

**Table 3**. Processes, principles, and pitfalls in the implementation phase of developing an educator-technology partnership.

| Implementation Phase | | |
|---|---|---|
| **Processes** | **Principles** | **Pitfalls** |
| Capacity Building<br>Ensure that the organization is able to effectively use the intervention (training, technology, infrastructure, workforce, budgets, policies/procedures, etc.).<br><br>Incremental Testing<br>Conduct small-scale tests and observe results to scale up progressively. This includes the intervention itself and also the capacity building efforts.<br><br>Adapting and Scaling-Up<br>Incorporate lessons learned into adapting the team, the theory, and the solution. Perform capacity building and incremental testing at a larger scale with each test. | Thorough Inquiry<br>Small-scale tests further explore organizational capacity, barriers, and various effects of the intervention.<br><br>Teamwork and Communication<br>Early successes are recorded and communicated to local stakeholders. Local stakeholders collaborate with each other on how to capitalize on early success and adapt to early setbacks.<br><br>Iterative Revision<br>Unexpected outcomes are not a problem, but an opportunity to make refinements to the theory of change and the overall plan for the project. | Exceeding organizational capacity<br><br>Insufficient resources allocated to testing and revision<br><br>Failure to adapt to early setbacks<br><br>Failure to communicate early successes |

## Capacity Building

Although the project should utilize the organization's existing strengths, advantages, and resources, implementing new human-technology partnerships will require building the capacity of the organization to effectively use the new intervention. The policies and procedures of the organization may need to be updated. Depending on the type of technology, electrical or telecommunications infrastructure may need to be augmented. The LAUSD and the OLPC caused problems by exceeding network capabilities. When interactive whiteboards were being introduced into many classrooms, they often failed not only because the schools lacked the technical support to make them function properly, but also the teachers were not trained on how these devices could augment their teaching practices. Without the proper technical support and training, these devices served as a sophisticated, error-prone means of teaching just the way they had taught before. The capacity building efforts of implementing a new educator-technology partnership will almost always involve building human capacities. This is where the support of organizational leadership is key. Teachers will need support from their leadership if they are expected to take on additional training, workload, or responsibilities. Whenever possible, limit these capacity building efforts to the scale of the next upcoming round of testing. Rebuilding infrastructure is expensive, and repeatedly retraining teachers is exhausting. Expect that small scale tests will lead to new insights into subsequent rounds of capacity building.

## Incremental Testing

There will inevitably be some unforeseen setbacks and some early successes. Whenever possible, do preliminary tests outside of classes. When moving to real-world tests in an active classroom, begin with small-scale tests in low-stakes settings. Unforeseen setbacks in an elective course taught by a seasoned instructor will have fewer negative consequences than in some other settings. Be considerate in how to frame this choice to the students; no one likes being guinea pigs in someone else's experiment. While testing the new intervention in a setting that won't cause too much harm if it goes wrong, test it in a way that will highlight potential failure mechanisms; thereby guiding essential revisions. Exposing these setbacks will enable revisions to all components of the implementation plan, including but not limited to: the features of the intervention, teamwork and communication strategies, measurement and monitoring approaches, understanding of stakeholder needs, understanding of organizational capacity and barriers, implementation goals and timelines, and quality assurance and stability mechanisms. This requires both a rigorous genuine attitude of seeking to learn, and a rigorous approach to measurement and monitoring. Including skeptics and naysayers in the team building processes will pay off here. This process has much in common with scientific research. The initial plans function as a hypothesis. A good hypothesis is testable. A good test is capable of falsifying the hypothesis. Unexpected findings lead to careful revisions of the theory. This updated theory generates a new hypothesis, or in other words, a new plan. This is why both qualitative and quantitative measurements are useful when monitoring the effects of incremental tests. Quantitative measurements often support or falsify existing factors in the Theory of Change, while qualitative methods are useful for finding missing factors.

In well-planned implementations, early attempts will also generate preliminary successes which should be communicated to a variety of stakeholders. Including leaders in the team-building process will pay off here. In many school settings, local leadership can have a powerful effect by offering meaningful support in the face of setbacks and spreading word of notable successes.

## Adapting and Scaling Up

The low-stakes settings appropriate for initial tests can sometimes focus on students with fewer challenges and obstacles than other students. When scaling up to larger tests, there are also issues with organizational structure and policy that may not come to light in small-scale tests. So, be aware that there is still a need to look out for unforeseen setbacks and new failure mechanisms with each subsequent expansion of the implementation. This is why the implementation process is incremental and iterative. The early planning phases should have identified the major factors at play, but early tests may justify major revision to the central premise of the project. As in previous phases, it is better to do more iterations on these processes than it is to rush ahead to the next phase prematurely.

# Key Considerations in the Sustainment Phase

Throughout the project, the new intervention may function as an appendage to the regular operations of the organization. The project is not over until the new intervention is incorporated into these regular operations. During the preparation phase, the implementation team should have already considered how the organization conducts regular quality assurance of its existing practices. It should have also considered how the organization has achieved stability of its existing practices. At this point, the Theory of Change is simplified to focus on the most crucial factors of how the new intervention interacts with the organizational context. This way, those factors can be intentionally designed into the schools policies, training programs, culture, and physical infrastructure. Table 4 shows important processes, principles, and potential pitfalls to consider in the sustainment phase of a new technology intervention in education.

**Table 4**. Processes, principles, and pitfalls in the sustainment phase of developing an educator-technology partnership.

| Sustainment Phase | | |
|---|---|---|
| **Processes** | **Principles** | **Pitfalls** |
| Stabilization<br>Incorporate the intervention into the regular operations of the organization, including stability mechanisms and quality assurance processes.<br><br>Phasing-Out External Support<br>Gradually allow the local organization to take full responsibility for using the new intervention. | Thorough Inquiry<br>Removing external support is an opportunity to expose any ongoing challenges, especially any necessary stability mechanisms and quality assurance processes.<br><br>Teamwork and Communication<br>Each component of the school organization has a clear understanding of their role within the new intervention. Once the school organization is managing the stability and | Lack of clarity regarding which organizational roles conduct ongoing quality assurance.<br><br>Failure to monitor progress while phasing out external support. |

| | quality assurance of the intervention, the local implementation team may dissolve.<br><br>Iterative Revision<br>The organization's ordinary practices for continued improvement and quality assurance may need to be adapted in response to ongoing challenges and opportunities. | |
|---|---|---|

## Stabilization

Local organizational members must sustain the intervention with no outside support. Few, if any, interventions provide value to the organization when used haphazardly. Each school will have their own mechanisms for conducting quality assurances of their internal practices. The ongoing quality assurance processes should support and enhance the causal mechanisms that make the intervention work. Thus, the Theory of Change must include those causal mechanisms and be simplified to focus on the most crucial factors. For example, if technology is implemented in a university classroom to enable interactive engagement pedagogy (i.e., active learning), and if the university uses end-of-semester feedback forms as its main quality assurance process, these feedback forms should allow students to comment on the level of interactive engagement in class. The long-term success of the intervention depends on local stakeholders continuing to conduct quality assurance of the intervention as a component of their routine organizational processes. Every organization has internal mechanisms for keeping certain features stable over time. This may include physical infrastructure, formal policies, teacher training programs, and organizational culture. From the earliest stages of the implementation project, the project should have begun by making partners of local stakeholders. If so, the project will have a major advantage at this final stage.

## Phasing-Out External Support

The implementation is not complete if the organization requires external support to train people to use the intervention, conduct ongoing quality assurance, maintain necessary infrastructure, or secure funding. Think of phasing out external support as another form of testing in that it is an opportunity to make observations, and when necessary, corresponding revisions. It may be that previous testing and adaptation has worked out all the bugs, or it could be that removing external support exposes new issues. Depending on the level of ongoing challenges, the sustainment phase may be as iterative as the previous three phases, or it could be mostly linear. If phasing out support exposes new challenges that require backtracking to additional stabilization, then it is better to backtrack and iterate than to push ahead.

# Protocols for Implementing Successful Educator-Technology Partnerships

Now that we have described the core principles, processes, and pitfalls of implementing educator-technology partnerships, we offer a set of user-friendly protocols for planning and conducting an implementation project. These include the major milestones that must be achieved before advancing from one phase to the next, the key indicators that these milestones have been fully achieved, and instructions for using SWOT analysis as an aid in brainstorming and strategic planning. Finally, we will characterize a mindset for working through these protocols most effectively.

## Milestones and Readiness Indicators

Implementing new human-technology partnerships, or any new practice, is never easy considering the level of planning, attention to detail, and responsiveness to setbacks required. Though the process is unavoidably difficult, it can be broken down into a series of achievable milestones. The goal is that eventually the new educator-technology partnership will no longer be an intervention that functions as an appendage to the organization, but is fully assimilated into the regular day-to-day operations. This level of assimilation can only be achieved if the new human-technology partnership leverages the unique advantages and overcomes the unique barriers in the local context. Revealing these key advantages and barriers is an iterative process of meticulous testing and revision. Meticulous, iterative testing and revision requires a revisable strategic plan. Such a careful plan can only be developed once a good intervention is selected, and selecting a good intervention depends on developing a thorough understanding of students' most important needs. Many failed implementations of educator-technology partnerships simply never achieved one or more of these basic milestones: the LAUSD did not consider how iPads would meet the needs of students or teachers, the OLPC project did not revise their plan after small-scale tests revealed unforeseen challenges, and many flipped classroom initiatives never leveraged the stability mechanisms within their university departments. In Table 5, we outline four major milestones corresponding to the four EPIS phases where each major milestone must be achieved before moving on to the next phase. Since moving forward with an implementation project without achieving each milestone is detrimental for success, we suggest key indicators of readiness to move forward.

**Table 5**. Major milestones in each of the four EPIS phases, and indicators of readiness to move to the next phase.

| Exploration Phase | Preparation Phase | Implementation Phase | Sustainment Phase |
|---|---|---|---|
| Major Milestones ||||
| **Selected an intervention that meets local needs**<br>○ Collaborated | **Developed a revisable strategic plan**<br>○ Built a team | **Exposed unforeseen challenges and iteratively refined the plan** | **Incorporated the new intervention into the regular operations of the** |
| with local stakeholders to outline local needs<br>○ Brainstormed multiple possible interventions to meet those needs<br>○ Made preliminary revision to the selected intervention to fit the context | representing all levels of the school organization<br>○ Mapped out the advantages and disadvantages within the organization (i.e. a Theory of Change [ToC])<br>○ Planned for the the implementation and sustainment phases | ○ Built the capacity of the organization to use the intervention<br>○ Conducted small-scale tests<br>○ Openly communicated successes and setbacks<br>○ Adapted the plan, revised the intervention, and scaled-up tests | **organization**<br>○ Imposed stability mechanisms and quality assurance processes<br>○ Phased out external support |
| Readiness Indicators | | | |
| ○ New challenges and opportunities were identified during the needs assessment<br>○ Local stakeholders made specific contributions to the needs assessment<br>○ Conflicting needs were identified during the needs assessment<br>○ Local stakeholders made specific contributions to the intervention fit assessment | ○ Local stakeholders from all levels of the organization have clear roles in the project<br>○ School leadership is ready to offer support when setbacks occur<br>○ ToC includes factors from the intervention, the organization, and the environment in which the organization operates<br>○ ToC includes the causal mechanisms that makes the intervention work<br>○ ToC includes stability mechanisms and quality assurance processes that may be leveraged in the late stages of the project<br>○ Local stakeholders made specific contributions to the capacity and barriers assessment<br>○ The plan for | ○ Unforeseen complications exposed via small-scale testing led to specific revisions to the intervention, to capacity-building strategies, and the ToC<br>○ Local stakeholders made specific contributions to these revisions | ○ The new intervention does not strain the capacity of the organization in terms of budgets, workloads, training, or infrastructure<br>○ Local stakeholders at all levels of the organization have a shared understanding of their role using and maintaining the new intervention for the long-term. |

| | monitoring and measurement includes both qualitative and quantitative measurements<br>○ The plan explicitly allocates time and resources for testing and revision | | |
|---|---|---|---|

## SWOT Analysis

Working through these steps of an implementation project involves intensive strategic planning, especially at the key milestones of selecting an intervention, and determining how to make that intervention a long-term success. Any educational organization will present a complex web of interacting advantages and disadvantages that must be understood in detail to achieve these key milestones. Some strategic planning tools which represent this inherent complexity are themselves so complex that they are difficult to use ([Powell et al., 2015](#)). SWOT analysis, on the other hand, is a strategic planning tool with a simple organizational scheme. Its purpose is to judge the feasibility of a prospective project by considering advantages and disadvantages for both internal and external sources ([Namugenyi et al., 2019](#)). Strengths, Weaknesses, Opportunities, and Threats (SWOT) are organized into a table. Despite being independently developed, the following recommendations for conducting a SWOT analysis ([Madsen, 2016](#); [Namugenyi et al., 2019](#)) are remarkably similar to the process described by Implementation Science:

1. Brainstorm factors to fill in the SWOT table in collaboration with all levels of the organization, and with outside sources when needed.
2. Carefully outline interactions among these factors and prioritize the most critical factors based on the potential impact on the project.
3. Develop strategies to overcome barriers and leverage advantages.
4. Adaptively implement the selected strategies with ongoing monitoring of progress, clear communication, and appropriate allocation of resources.

Though these recommendations are a direct analog of the processes described by Implementation Science, there is one important difference. In SWOT analysis, the step of developing strategies is framed as a creative problem-solving process rather than a formulaic process ([Jarzabkowski et al., 2007](#); [Whittington, 2003](#)). This fills in a notable gap in Implementation Science, but also benefits from small modifications from Implementation Science. Whereas Project Planning Research considers external and internal influences, Implementation Science additionally considers the unique advantages and disadvantages of the new intervention itself. To capture this, we adjust the SWOT table from its traditional two-by-two format to a two-by-three format, as shown in Table 6.

**Table 6**. Example of a SWOT analysis based on the One Laptop Per Child project. In this case, low-income nations are the external environment, the school system within those nations is the organization, and the provision of laptops for children is the intervention.

|  | Advantages | Disadvantages |
|---|---|---|
| **External Environment**<br><br>Cultural<br>Political<br>Socioeconomic<br>Legal<br>Etc. | - Premise is appealing to funders in high-income nations | - Bigger problems in low-income nations than low computer literacy<br><br>- Unreliable or non-existant electricity and telecommunications infrastructure<br><br>- Low income in the Global South reduces the relative cost effectiveness of technological solutions over human-centered solutions such as creating schools and libraries<br><br>- Minimal interest in computers |
| **Organization**<br><br>Leadership Structure<br>Staffing<br>Training and Expertise<br>Emotional Factors<br>Policy and Procedures<br>Technological Capacity<br>Budgets<br>Etc. |  | - Minimal interest in the role of computers in education<br><br>- Minimal local capacity to provide tech support<br><br>- Minimal capacity to provide training for students on how to use the units |
| **Intervention**<br><br>Scalability<br>Adaptability<br>Context Dependence<br>Cost<br>Technological Constraints<br>Required Expertise<br>Causal Mechanisms<br>Etc. | -Cost per unit decreases with increasing scale<br><br>-Cheap for funders in high-income nations<br><br>-Adaptable for users without literacy in their own language<br><br>-Units are durable | - Manufacturing and logistics are insufficient for large-scale production and distribution<br><br>- Cost per unit is extremely high relative to the yearly income of the users<br><br>- Requires reliable electricity and telecommunications infrastructure<br><br>- Requires training to integrate into schools<br><br>- Requires ongoing tech |

| | | | support to use and maintain the units |
|---|---|---|---|

We provide an example of this adapted SWOT analysis for the OLPC project in Table 6. With the benefit of hindsight, this project can serve as an example of how to consider stakeholder needs and important contextual constraints in an implementation project. The SWOT analysis helps identify various interrelated challenges and opportunities for this example project. On one hand, simply building schools and libraries could have a greater impact at less cost. On the other hand, the laptops would target the specific problem of assisting users without literacy in their own language. The SWOT analysis leads to a possible strategy of focusing on schools and libraries in most cases, and using the laptops in cases where additional support for developing literacy is needed and there is sufficient local infrastructure to support the laptops. In these special cases, the project would need to provide external technical support. The plan for this strategy should involve piloting the laptops and carefully monitoring the level of external support required, and measuring the additional benefit beyond simple schools and libraries.

This strategic analysis tool cannot compensate for failure to fully explore local needs prior to selecting an intervention. Neither can this tool force planners to attend to the constraints faced by the stakeholders; it merely provides an *opportunity* to attend to those constraints (Madsen, 2016; Namugenyi et al., 2019). This example also demonstrates the value of considering multiple possible solutions to meet stakeholder needs amid their constraints. SWOT analysis is unlikely to rescue a project that moves ahead with a single pre-selected intervention based on a poor understanding of the stakeholders needs and constraints. Thus, we recommend first performing an assessment of local needs, then considering multiple possible interventions, and then conducting a SWOT analysis to help explore important constraints. The SWOT table should be filled out in moderate detail for each potential intervention being considered in the exploration phase, and it should be filled out in great detail for the selected intervention in the preparation phase. In the preparation phase, the SWOT table combines the findings of the needs assessment, the intervention fit assessment, and the barriers and capacity assessment. Craft a detailed Theory of Change by finding the interconnections between the factors in the SWOT table and prioritizing the strongest influences on the project.

## Think Like an Expert

Without the proper mindset, it is possible to work through these protocols in a perfunctory manner and generate underwhelming results. When designing a project to implement a new educator-technology partnership; researchers, educators, and technology developers often reproduce common errors made by novice designers. Novices tend to make the same mistakes in any field. Thus, anyone without vast experience working to implement new technology and practices into educational organizations would be prudent to consider themself a novice. The purpose of this mindset is to be on alert for common pitfalls, and to consciously mimic an expert-like approach.

Notable failures like the OLPC laptops and LAUSD iPads are not surprising considering that people often do not continue generating ideas once they have even one solution in mind (Binz & Schulz, 2023; Jansson & Smith, 1991). Though experts also experience the phenomenon of fixation on initial ideas, they tend to utilize strategies to overcome this problematic mental barrier whereas novices tend to make choices which exacerbate it (Crismond & Adams, 2013; Atman et al., 2013). Specifically, experts will conjecture multiple solutions as a way of exploring constraints in the problem (Cross, 2004). Whereas in mechanical engineering, the relevant constraints include materials, manufacturing, and maintenance; in education, these constraints include teacher training, classroom infrastructure, limited workforce, etc. After exploring constraints, experts will then re-analyze the problem and represent it in a way that is entirely independent of how they frame the solutions (Fricke, 1999). Once the problem is defined, experts will search for pre-existing solutions (Crismond & Adams, 2013). They will expect errors in their initial design and allocate sufficient time and resources for testing and revision. Novices tend to analyze the problem with a particular solution in mind, and frame the problem in terms of their initial solution (Fricke, 1999; Flynn, 2020). Framing the problem in this novice manner contributes to the common pitfall of failure to consider constraints, including potential failure modes and critical needs of the client or end-user (Loweth et al., 2020). Novices will therefore tend to be overconfident in the effectiveness and uniqueness of their solution (Atman et al., 2013). This can lead to "reinventing the wheel" rather than searching for existing solutions. It also leads to planning for rigid timelines that lack sufficient time or resources allocated to testing and revision (Flynn, 2020; Crismond & Adams, 2013).

Poor team dynamics are another major contributor to failure in nearly any field (Leifer, 1998; Whitcomb & Whitcomb 2013; Maier et al, 2021). Often because novices do not make intentional strategic choices about their approach to teamwork and communication. This results in poorly defined roles among team members (Jantzer et al, 2020). Experts include the client or end-user in the early collaborative process of framing the problem and outlining constraints. In contrast, novices do not frame the problem in collaboration with the end-user, leading to a poorer understanding of user needs (Loweth et al., 2020; Flynn, 2020; Atman et al., 2013). By applying so much rigor to fully understand a problem, experts are able to construct a testable *predictive framework* for the project at hand. This predictive framework is equivalent to a Theory of Change (Rapport et al. 2022; Breuer et al., 2016) in that it is a detailed map of the strongest factors influencing the outcomes, and is subjected to iterative testing and revision throughout the course of the project.

The essence of the expert mindset is to anticipate revision, and thus approach any project in a way that is inherently inquisitive, collaborative, and iterative (Crismond & Adams, 2013; Flynn, 2020). The value of the expert strategies and the consequences of the novice mistakes are magnified when working on ill-structured problems that have any combination of conflicting definitions of success, multiple stakeholders, complex causal relationships, ambiguous data, and intricate human factors (Simon, 1973). Developing human-technology partnerships in education will always be an ill-structured problem. Although these partnerships can be successfully implemented using some of the most basic principles of Implementation Science, it is important to appreciate the level of careful planning required, and adopt a mindset suited to the challenge.

While it is an appealing notion that educators will spontaneously adopt any new technology that eases the burdensome tasks they must perform, we recommend an alternative outlook. New technology, just like non-technological improvements, are far more likely to achieve long-term adoption if they are developed in collaboration with the end-user and implemented via carefully-planned processes. With a mindset that implementing new technology into educational organizations is an ill-structured problem, educators and technology developers alike may utilize the problem-solving strategies they already know. We offer the EPIS Framework and SWOT Analysis (each slightly modified) as complementary tools to find long-term success in meeting the needs of teachers and students.

# References

Aarons, Gregory A., et al. "Advancing a Conceptual Model of Evidence-Based Practice Implementation in Public Service Sectors." *Administration and Policy in Mental Health and Mental Health Services Research*, vol. 38, no. 1, 14 Dec. 2011, pp. 4–23,

Allen, Scott E, and René F Kizilcec. "A Systemic Model of Academic (Mis)Conduct to Curb Cheating in Higher Education." *Higher Education*, 19 July 2023, https://doi.org/10.1007/s10734-023-01077-x.

Ames, Morgan G. *The Charisma Machine : The Life, Death, and Legacy of One Laptop per Child*. Cambridge, Massachusetts, The Mit Press, 2019.

Atman, Cynthia J., et al. "Engineering Design Processes: A Comparison of Students and Expert Practitioners." *Journal of Engineering Education*, vol. 96, no. 4, Oct. 2007, pp. 359–379, https://doi.org/10.1002/j.2168-9830.2007.tb00945.x. Accessed 5 Dec. 2021.

Binz, Marcel, and Eric Schulz. "Reconstructing the Einstellung Effect." *Computational Brain & Behavior*, vol. 6, no. 3, 14 Dec. 2022, pp. 526–542, https://doi.org/10.1007/s42113-022-00161-2.

Breuer, Erica, et al. "Using Theory of Change to Design and Evaluate Public Health Interventions: A Systematic Review." *Implementation Science*, vol. 11, no. 1, Dec. 2015.

Crismond, David P. , and Robin S. Adams. "The Informed Design Teaching and Learning Matrix." *Journal of Engineering Education*, vol. 101, no. 4, Oct. 2012, pp. 738–797, https://doi.org/10.1002/j.2168-9830.2012.tb01127.x.

Cross, Nigel. "Expertise in Design: An Overview." *Design Studies*, vol. 25, no. 5, Sept. 2004, pp. 427–441, https://doi.org/10.1016/j.destud.2004.06.002.

Cuban, Larry. *Oversold and Underused : Computers in the Classroom*. Cambridge, Mass. ; London, Harvard University Press, 2001.

Damschroder, Laura J., et al. "The Updated Consolidated Framework for Implementation Research Based on User Feedback." *Implementation Science*, vol. 17, no. 1, 29 Oct. 2022, pp. 1–16, https://doi.org/10.1186/s13012-022-01245-0.

Dancy, Melissa, and Charles Henderson. "Pedagogical Practices and Instructional Change of Physics Faculty." *American Journal of Physics*, vol. 78, no. 10, Oct. 2010, pp. 1056–1063, https://doi.org/10.1119/1.3446763.

Daniel, John. "Making Sense of MOOCs: Musings in a Maze of Myth, Paradox and Possibility." *Journal of Interactive Media in Education*, vol. 2012, no. 3, 13 Dec. 2012, p. 18, https://doi.org/10.5334/2012-18.

Flynn, Michael . *Assessing Expertise in Mechanical Engineering Design*. 2020.

Freeman, S., et al. "Active Learning Increases Student Performance in Science, Engineering, and Mathematics." *Proceedings of the National Academy of Sciences*, vol. 111, no. 23, 12 May 2014, pp. 8410–8415, https://doi.org/10.1073/pnas.1319030111.

Fricke, Gerd. "Successful Approaches in Dealing with Differently Precise Design Problems." *Design Studies*, vol. 20, no. 5, Sept. 1999, pp. 417–429, https://doi.org/10.1016/s0142-694x(99)00018-6.

Gurl, Emet. "SWOT ANALYSIS: A THEORETICAL REVIEW." 11 Aug. 2017, demo.dspacedirect.org/items/a94ef210-25e1-4399-b93e-54930f2ba37f, http://dx.doi.org/10.17719/jisr.2017.1832.

Jansson, David G., and Steven M. Smith. "Design Fixation." *Design Studies*, vol. 12, no. 1, Jan. 1991, pp. 3–11, https://doi.org/10.1016/0142-694x(91)90003-f.

Jantzer, Michael, et al. "Roles in Engineering." *Springer EBooks*, 1 Jan. 2020, pp. 133–138, https://doi.org/10.1007/978-3-662-60384-0_18.

Jarzabkowski, Paula, et al. "Strategizing: The Challenges of a Practice Perspective." *Human Relations*, vol. 60, no. 1, Jan. 2007, pp. 5–27, journals.sagepub.com/doi/abs/10.1177/0018726707075703, https://doi.org/10.1177/0018726707075703.

Kraemer, Kenneth L., et al. "One Laptop per Child: Vision vs. Reality." *Communications of the ACM*, vol. 52, no. 6, 1 June 2009, p. 66, https://doi.org/10.1145/1516046.1516063.

Leifer, Larry. "Design-Team Performance: Metrics and the Impact of Technology." *Springer EBooks*, 1 Jan. 1998, pp. 297–319, https://doi.org/10.1007/978-94-011-4850-4_14.

Lin, Hong. "Implementing LargeScale Mobile Device Initiatives in Schools and Institutions." *Emerging Technologies for STEAM Education: Full STEAM Ahead*, edited by Xun Ge et al., Cham, Springer International Publishing, 2015, pp. 179–198, doi.org/10.1007/9783319025735_10.

Lovett, Marsha C., et al. *How Learning Works: Eight Research-Based Principles for Smart Teaching*. *Google Books*, John Wiley & Sons, 14 Mar. 2023.

Loweth, Robert P, et al. "Novice Designers' Approaches to Justifying User Requirements and Engineering Specifications." *ASME 2020 International Design Engineering Technical Conferences and Computers and Information in Engineering Conference*, 17 Aug. 2020, https://doi.org/10.1115/detc2020-22163.

Madsen, Dag Øivind. "SWOT Analysis: A Management Fashion Perspective." *Ssrn.com*, 2016, papers.ssrn.com/sol3/papers.cfm?abstract_id=2615722.

Maier, Anja M., et al. "Factors Influencing Communication in Collaborative Design." *Journal of Engineering Design*, 27 July 2021, pp. 1–32, https://doi.org/10.1080/09544828.2021.1954146.

Moullin, Joanna C., et al. "Systematic Review of the Exploration, Preparation, Implementation, Sustainment (EPIS) Framework." *Implementation Science*, vol. 14, no. 1, 5 Jan. 2019, implementationscience.biomedcentral.com/articles/10.1186/s13012-018-0842-6, https://doi.org/10.1186/s13012-018-0842-6.

Moullin, Joanna C., et al. "Ten Recommendations for Using Implementation Frameworks in Research and Practice." *Implementation Science Communications*, vol. 1, no. 1, 30 Apr. 2020, https://doi.org/10.1186/s43058-020-00023-7.

Perry, Cynthia K., et al. "Specifying and Comparing Implementation Strategies across Seven Large Implementation Interventions: A Practical Application of Theory." *Implementation Science*, vol. 14, no. 1, 21 Mar. 2019, https://doi.org/10.1186/s13012-019-0876-4.

Rapport, Frances, et al. "Too Much Theory and Not Enough Practice? The Challenge of Implementation Science Application in Healthcare Practice." *Journal of Evaluation in Clinical Practice*, 15 July 2021, https://doi.org/10.1111/jep.13600.

Reich, Justin. *Failure to Disrupt: Why Technology Alone Can't Transform Education*. *Google Books*, Harvard University Press, 15 Sept. 2020.

Ross, Steven M. "Technology Infusion in K-12 Classrooms: A Retrospective Look at Three Decades of Challenges and Advancements in Research and Practice." *Educational Technology Research and Development*, vol. 68, no. 5, 12 Mar. 2020, https://doi.org/10.1007/s11423-020-09756-7.

Schwartz, Daniel, et al. *The Abcs of How We Learn : 26 Scientifically Proven Approaches, How They Work, and When to Use Them.* W W Norton & Co Inc, 2016.

Simon, Herbert A. "The Structure of Ill Structured Problems." *Artificial Intelligence*, vol. 4, no. 3-4, 1973, pp. 181–201, https://doi.org/10.1016/0004-3702(73)90011-8.

Whitcomb, Clifford A, and Leslie E Whitcomb. *Effective Interpersonal and Team Communication Skills for Engineers*. Hoboken, New Jersey, Ieee Press ; Hoboken, New Jersey, 2013.

Whittington, Richard. "The Work of Strategizing and Organizing: For a Practice Perspective." *Strategic Organization*, vol. 1, no. 1, Feb. 2003, pp. 117–125, https://doi.org/10.1177/147612700311006.

Wieman, C E. *Improving How Universities Teach Science : Lessons from the Science Education Initiative*. Cambridge, Massachusetts, Harvard University Press, 2017.

WIRED Staff. "What Schools Must Learn from LA's IPad Debacle." *WIRED*, WIRED, 8 May 2015, www.wired.com/2015/05/los-angeles-edtech/.